\newcommand{\apj}{ApJ}
\newcommand{\aap}{A\&A}
\newcommand{\etal}{et al.}

\newcommand{\lee}{{\em left}}
\newcommand{\rii}{{\em right}}
\newcommand{\topp}{{\em top}}
\newcommand{\bott}{{\em bottom}}

\newcommand{\rms}{{\emph{rms}}}

\newcommand{\dss}{$\delta$~Scuti}
\newcommand{\epscep}{$\epsilon$~Cep}
\newcommand{\epscepp}{$\epsilon$~Cephei}

\newcommand{\templogg}{{\sc templogg}}

\newcommand{\hipp}{{\sc hipparcos}}
\newcommand{\str}{Str\"omgren}
\newcommand{\period}{{\sc period04}}
\newcommand{\wire}{{\sc wire}}
\newcommand{\osn}{{\sc osn}}
\newcommand{\corot}{{\sc corot}}
\newcommand{\most}{{\sc most}}

\newcommand{\kepler}{{\sc kepler}}
\newcommand{\simbad}{{\sc simbad}}

\newcommand{\eg}{e.g.}
\newcommand{\cf}{cf.}
\newcommand{\ie}{i.e.}
\newcommand{\vs}{vs.}

\newcommand{\teff}{$T_{\rm eff}$}
\newcommand{\logg}{$\log g$}
\newcommand{\feh}{[Fe/H]}
\newcommand{\vsini}{$v \sin i$}
\newcommand{\kms}{km\,s$^{-1}$}
\newcommand{\hbeta}{H$_\beta$}
\newcommand{\panel}{{\em panel}}
\newcommand{\panels}{{\em panels}}
\newcommand{\cday}{c/day}

\documentclass[twocolumn]{aa}

\usepackage{txfonts}
\usepackage{graphicx}
\usepackage{natbib}
\bibpunct[, ]{(}{)}{;}{a}{,}{,}
 
\usepackage{graphicx}
\usepackage{txfonts}

\begin{document}

\title{Asteroseismology with the \wire\ satellite}
\subtitle{I. Combining Ground- and Space-based Photometry of the $\delta$~Scuti~Star \epscepp}

   \author{H.~Bruntt
           \inst{1,2}
            \and
          J.C.~Su\'arez
           \inst{3,4}
\thanks{Associate researcher at l'Observatoire de Paris.}
            \and
          T.R.~Bedding\inst{2}
            \and
          D.L.~Buzasi
           \inst{5}
            \and
          A.~Moya
           \inst{4}
            \and
P.J.\ Amado\inst{6,3}
\and
S.\ Mart\'{\i}n-Ruiz\inst{3}
\and
          R.~Garrido
           \inst{3}
            \and
P.\ L\'opez de Coca\inst{3}
            \and
          A.~Rolland
           \inst{3}
            \and
          V.~Costa
           \inst{3}
            \and
          I.~Olivares
           \inst{3}
            \and
          J.M.~Garc\'{\i}a-Pelayo
           \inst{3}
          }

   \offprints{H.~Bruntt}

   \institute{Niels Bohr Institute, Juliane Maries Vej 30, 
              University of Copenhagen, Denmark
              \email{bruntt@physics.usyd.edu.au}
         \and
             {School of Physics A28, University of Sydney, 2006 NSW, Australia}
              \email{t.bedding@physics.usyd.edu.au}
         \and
             Instituto de Astrof\'{\i}sica de Andaluc\'{\i}a, CSIC, CP3004,
             Granada, Spain             
             \email{jcsuarez@iaa.es} 
         \and
             Observatoire de Paris, LESIA, UMR 8109, Meudon, France
             \email{andy.moya@obspm.fr}
         \and 
             US Air Force Academy, Department of Physics, CO, USA,
             \email{derek.buzasi@usafa.af.mil}
         \and
            Universidad de Granada, Departamento F\'{\i}sica Te\'orica y del Cosmos, 
            Campus Fuentenueva, Granada, Spain
         }

   \date{Received xxx; accepted yyy}

  \abstract
   {}
   {We have analysed ground-based multi-colour \str\ photometry and 
single-filter photometry from the star tracker on the \wire\ satellite 
of the \dss\ star \epscepp.}
   {The ground-based data~set consists of 16 nights of data collected over 164 days, 
while the satellite data are nearly continuous coverage of the star 
during 14 days. The spectral window and noise level of the satellite data are
superior to the ground-based data and this data~set is used to locate the 
frequencies. However, we can use the ground-based data to
improve the accuracy of the frequencies due to the much longer time baseline. }
   {We detect 26 oscillation frequencies in the \wire\ data~set, 
but only some of these can be seen clearly in the ground-based data.
We have used the multi-colour ground-based photometry
to determine amplitude and phase differences in the \str\ $b-y$ colour 
and the $y$ filter in an attempt to identify the radial degree of the oscillation frequencies.
We conclude that the accuracies of the amplitudes 
and phases are not sufficient 
to constrain theoretical models of \epscep.
We find no evidence for rotational splitting or the large separation 
among the frequencies detected in the \wire\ data~set.}
   {To be able to identify oscillation frequencies in \dss\ stars with the method we have 
applied, it is crucial to obtain more complete coverage from multi-site campaigns 
with a long time baseline and in multiple filters. 
This is important when planning photometric 
and spectroscopic ground-based support for future satellite missions like \corot\ and \kepler.}


   \maketitle
%

%
     \begin{figure*}
     \centering
      \includegraphics[width=17.8cm]{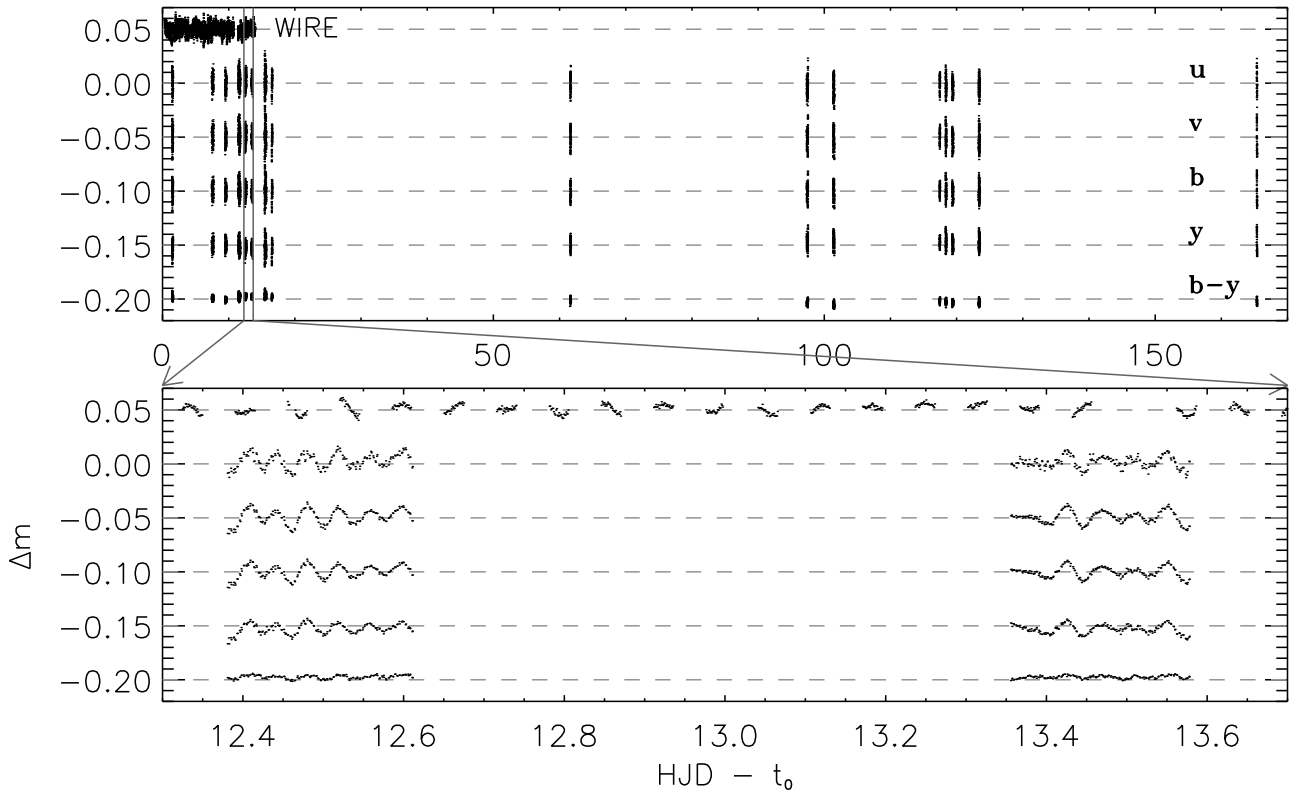} 
     \caption{Each \panel\ shows the light curve of \epscep\ from \wire\ and \osn\ in the 
      $uvby$ filters and $b-y$.
      The \topp\ \panel\ shows the complete set of light curves and 
      the \bott\ \panel\ shows the details of a small part.
      Note that the \osn\ data were collected about 2.9 years prior to the \wire\ data, 
      \ie\ the zero points $t_0$ are different (see main text).
                \label{fig:lc}}
      \end{figure*}

\section{Introduction}

\dss\ stars are main sequence population~I A- and F-type stars. 
They are found in the classical Cepheid instability strip
on the main sequence, have masses around twice solar, 
and temperatures around 7\,500~K. 
They have only very shallow
outer convection zones but their cores are fully convective. 
Most stars found in the instability strip are variable 
and most are multi-periodic, with
periods around 1--2 hours and amplitudes typically at the mmag level,
although some have amplitudes above 0.1 mag \citep{rodriguez00}.

To get a better understanding of \dss\ stars,
several multi-site ground-based campaigns have carried 
out extensive monitoring of selected targets.
For example, the Delta Scuti Network \citep{zima97, zima02}
has monitored FG~Virginis during several
seasons and 79 frequencies have been detected \citep{breger05}. In that
case, data from several ground-based observatories collected over 13 years
were combined, which allowed the detection of frequencies down to amplitudes 
of just 0.2 mmag in \str\ $y$. It is important to note that 
about 50 of the reported frequencies have amplitudes below 0.5 mmag. 
\citet{breger05} stated that typical
multi-site campaigns with a duration of 200 to 300\,h detect only 
5--10 frequencies. Since so few frequencies are detected, while
theory predicts a much higher number of excited modes, \citet{breger05} 
concluded that either longer ground-based photometric campaigns or
high-precision space-based campaigns are necessary. 

\citet{buzasi05} recently found Altair to be a low-amplitude
\dss\ star based on high-precision photometry from the star tracker on the
Wide-field InfraRed Explorer (\wire) satellite. They found only 
seven frequencies but three of these had amplitudes around 0.1 mmag.
The reason why so few frequencies were detected in Altair compared to
FG~Vir may be the much higher rotation rate of Altair, 
which has \vsini\ $=210\pm20$ \kms\ while FG~Vir has \vsini\ around $21$\,\kms\ \citep{mante02,mitter03}.
\citet{suarez05} have calculated models of Altair and used the
constraints from the frequencies detected by \citet{buzasi05}. 
\citet{suarez05} found that high rotational
velocity makes the interpretation of the frequencies difficult, due to
the limits of the applied second order perturbation theory
and effects of near degeneracy.
The \dss\ star treated in this paper has \vsini\ 
$\simeq90$ \kms\ and may present a simpler case. 
The studies of \citet{breger05} and \citet{buzasi05} both 
agree that high photometric precision and long temporal coverage is 
needed to fully explore the oscillation spectra of \dss\ stars. 

In the present study we have combined two quite different 
photometric data~sets for the multi-mode \dss\ star \epscepp\ (HD~211336). 
One data~set is from the \wire\ satellite covering 14 days
with very high signal-to-noise (S/N). The other data~set is 
single-site ground-based \str\ $uvby$ photometry from 
16 nights collected during 164 days, thus having gaps from 
a single day to several weeks. We find that by 
combining the very different properties of the data~sets
in terms of S/N and spectral window, we can measure the 
frequencies very accurately. 
We assess the uncertainties on the measured frequencies from
extensive simulations to see if mode identification is 
possible using amplitude ratios and phase differences in the
modes measured with the \str\ filters.

%
\subsection{Previous studies of \epscep\label{sec:previous}}

\epscep\ was first identified as a \dss\ star by \citet{breger66} who found
a single frequency at $f = 23.8\pm1.7$ c/day based on
only two nights of data. One night of $y$ band photometry 
by \citet{fesen73} showed an apparent change in amplitude, but
today we know that this is likely due to beating of frequencies.
Spectroscopy was first carried out during three nights by \citet{gray71}, 
who confirmed the period known from photometry.
The mean radial velocity amplitude measured on the three nights was $15\pm3$ \kms,
but this disagrees with upper limits of 
$\sim1$\,\kms\ found by \citet{kennelly99} and \citet{baade93}.

Line profile variations (LPVs) were detected by
\citet{baade93}. Based on only nine spectra they suggested that
the observed LPV could be explained by a $p$ mode with 
high azimuthal order $m\sim6-8$. More extensive monitoring
with high-resolution spectroscopy was done by \citet{kennelly99}.
They monitored \epscep\ in a multi-site campaign 
during eight nights. 
They detected a rich set of frequencies in the range 17--40 c/day 
with radial degrees $l=5-15$ using
two-dimensional Fourier analysis, but their results were preliminary.

\citet{costa03} monitored \epscep\ 
with \str\ filters but they concluded that the rich amplitude
spectrum could not be adequately resolved based on 
their single-site data~set. 
We will use this data~set in the present study.

\section{Observations\label{sec:obs}}

\subsection{Observations from the ground}

\epscepp\ (HD 211336) was monitored with simultaneous $uvby$ measurements
from Observatorio de Sierra Nevada (\osn) in 2001--2 on 16 nights during 
a period of 164 days \citep{costa03}. 
Eight nights of data were collected from 2001 August 9--25.
There is data from October 8 and two nights on 2001 November 13 and 17.
Four nights were obtained from 2001 December 3 to 9.
Finally, data from a single night was obtained on 2002 January 20.

In total \epscep\ was observed for
16 nights from \osn\ with typically 6--8 hours of observations each night.
A total of 2250 data points were collected.
After removal of the oscillations the \rms\ noise 
is 3.3 mmag in $u$ and 1.8 mmag in $v,b$ and $y$.
The complete light curve is shown in the \topp\ \panel\ in Fig.~\ref{fig:lc}
while details of the stellar oscillations are seen in the \bott\ \panel.
The zero point in time is $t_0 =  2\,452\,130$.

\subsection{Observations from space}

The Wide-field InfraRed Explorer (\wire) satellite mission was
designed to study star-burst galaxies in the infrared \citep{hacking99}.
Unfortunately,
the hydrogen which was to be used for cooling the main camera was
lost soon after launch. Since 1999 the 52\,mm star tracker on 
\wire\ has been used to monitor bright stars continuously
for one to six weeks \citep[see][]{bruntt06}. 

\epscep\ was observed with \wire\ from 2004 June 20 to July 4.
The raw data~set consists of around 600\,000 8x8 pixel 
CCD windows centered on the star with a time-sampling of 0.5\,s.
The data were reduced as 
described by \citet{bruntt05} and points taken within 15\,s were binned.
The resulting light curve has 25\,293 data points collected during 13.6 days
with three short gaps with durations of 0.2, 0.2, and 0.5 days. 
The complete light curve is shown in Fig.~\ref{fig:lc} where the zero point in time
is $t_0 =  2\,453\,175.5$.   
Note that the \wire\ observations 
started about 2.9 years after the \osn\ run. 


The \rms\ noise level in the \wire\ data~set after removal of the oscillations
is 1.7 mmag. To estimate the white noise component we calculated the noise in 
the amplitude spectrum at high frequencies 
($10\pm0.5$ mHz; the Nyquist frequency is 33.3 mHz). 
From this we found the noise level to be 12.3~ppm or 1.2 mmag per 15\,s bin.
Each \wire\ observation collects about $e_{\rm star}=10^5$ electrons, and after binning
every 30 data points (15\,s sampling), the theoretical Poissonian noise is
0.6\,mmag or more than a factor two lower than the actual observed noise level.
The higher noise level is due to the relatively high sky level during the \epscep\ run. 
We estimate the noise contribution from the background to be
$\sigma_{\rm AP}^2=n_{\rm pix} * (e_{\rm sky} + \sigma_{\rm ro}^2) / e_{\rm star}^2$
following eq.~31 in \cite{kjeldsen92}; here $n_{\rm pix}$ is the number of pixels
in the aperture, $\sigma_{\rm ro}$ is the readout noise,
while $e_{\rm star}$ and $e_{\rm sky}$ are the number of electrons 
from the star and sky background, respectively.
We use $n_{\rm pix}=12$, a gain of 15 $e^{-1}$/ADU, $e_{\rm sky}=420\pm220$\,ADU,
and a $\sigma_{\rm ro}=10$\,$e^{-1}$ to obtain $\sigma_{\rm AP}=2.7\pm0.7$\,mmag which is
comparable to the Poissionian noise $\sigma_{\rm count}=1/\sqrt{e_{\rm star}}=3.2$\,mmag.
This explains the relatively high noise level.

\begin{table}
\caption{Basic photometric indices for \epscepp. 
\label{tab:phot}}      
\centering
\begin{tabular}{rr|rrr|r}
   \hline\hline       
   $V$ & $B-V$ & $b-y$ & $m_1$ & $c_1$ & $H_\beta$ \\
   \hline
   4.19 & 0.28 & 0.171(2) & 0.192(4) & 0.784(4) & 2.761 \\
   \hline                    
   \hline                  
\end{tabular}
\end{table}

\section{Fundamental parameters of \epscep\label{sec:param}}

The basic photometric indices for \epscep\ are summarized in Table~\ref{tab:phot}.
The $V$ magnitude and $B-V$ colour are based on 12 measurements and 
are taken from Mermilliod's compilation
of Eggen's $UBV$ data (available through \simbad).
The \str\ indices are from \citet{hauck98} and are based on a combination of 
60 measurements, while \hbeta\ is based on 47 measurements.
The projected rotational velocity (\vsini) of \epscep\ is 105\,\kms\ according 
to 10 measurements from \citet{bernacca70}, while \citet{royer02} found 91\,\kms. 
The typical uncertainty on \vsini\ is 5\% \citep{royer02}, \ie\ $\sigma$(\vsini)\,$=5$\,\kms.

We used \templogg\ \citep{rogers95}
to determine the fundamental atmospheric parameters of 
\epscep\ and the results are 
\teff$=7340\pm50$ , \logg$=3.9\pm0.1$, 
\feh$=0.12\pm0.04$. This is consistent with the spectral type F0\,IV. 
We stress that the quoted uncertainties are based solely on the 
uncertainty on the photometric indices. Realistic uncertainties on \teff, \logg, and \feh\ are 
150~K, 0.2 dex, and 0.2 dex \citep{rogers95, kupka01}. 
Based on the \str\ $b-y$ and \hbeta\ indices
there is no significant interstellar reddening.

\citet{erspamer02,erspamer03} used an automated procedure 
to determine individual abundances of 
140 A-- and F--type stars and \epscep\ was included in their data~set.
They used Geneva photometry to fix 
\teff\ $=7244\pm150$~K and the \hipp\ parallax and evolution models
to find \logg\ $=4.02\pm0.2$, which both agree with our \str\ photometry
when using \templogg. The abundance analysis of \epscep\ yielded
\feh\ $=+0.08\pm0.10$, which is also in good agreement with \templogg.
The error estimate on \feh\ given here is based on the various 
contributions to the error budget, as discussed in detail 
by \citet{erspamer02}.

   \begin{figure}
   \centering
\includegraphics[width=8.8cm]{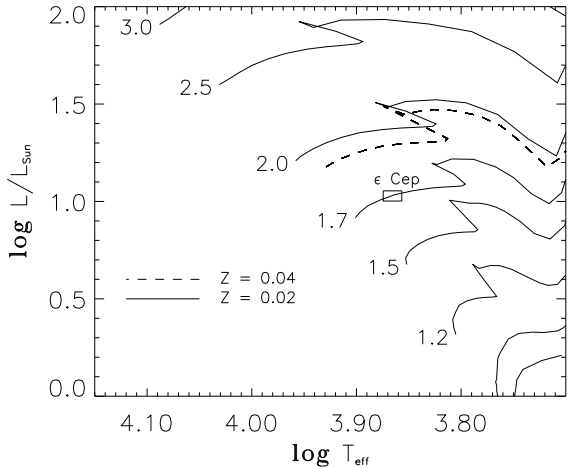}
 \caption{HR diagram with evolution tracks from \citet{lejeune} for
$Z=0.02$ (solar) and $Z=0.04$.
The 1\,$\sigma$ error box for \epscep\ is indicated and is
consistent with a slightly evolved star with mass $M/M_\odot=1.75\pm0.20$
for a metallicity of \feh\ $= 0.1$. 
\label{fig:hr}}
    \end{figure}

The location of \epscep\ in the Hertzsprung-Russell (HR) diagram is shown in
Fig.~\ref{fig:hr}. We show evolutionary tracks from \citet{lejeune} for 
solar metallicity ($Z=0.02$). The dashed track is for a metallicity
of twice the solar value for a mass $M/M_\odot = 2.0$.
To estimate the luminosity, we used the visual magnitude
$V = 4.19\pm0.03$ and the \hipp\ parallax 
of $38.9\pm0.5$ mas \citep{hipparcos97}. We found the bolometric 
correction (BC) by interpolation in the tables by \citet{bessell},
\ie\ BC\,$=0.04\pm0.02$. For the solar bolometric magnitude we used
$4.75\pm0.04$. Thus, we find $L/L_\odot = 10.7\pm0.6$ and 
adopt \teff\,$=7340\pm150$~K. The 1\,$\sigma$ error box is indicated
in Fig.~\ref{fig:hr}.

From the location of \epscep\ in the HR diagram relative 
to the evolutionary tracks from \citet{lejeune},
and adopting a metallicity \feh\,$=0.1\pm0.1$, we estimate the mass to be $M/M_\odot = 1.75\pm0.20$.

From the estimated mass, temperature, and parallax we can calculate
the surface gravity using 
$\log g_\pi = 4 [T_{\rm eff}] + [M] + 2 \log \pi + 0.4 (V + BC_V + 0.26) + 4.44$,
where $[T_{\rm eff}] = \log (T_{\rm eff}/T_{{\rm eff}\,\odot})$ and 
$[M] = \log (M/M_\odot)$. For \epscep\ we find 
$\log g_\pi = 4.0 \pm 0.1$ which agrees with \cite{erspamer03} and
the photometric calibration from \templogg.

%
%
\section{Time Series Analysis\label{sec:window}}
\subsection{Spectral windows\label{sec:window1}}

\begin{figure}
   \centering
    \includegraphics[width=8.8cm]{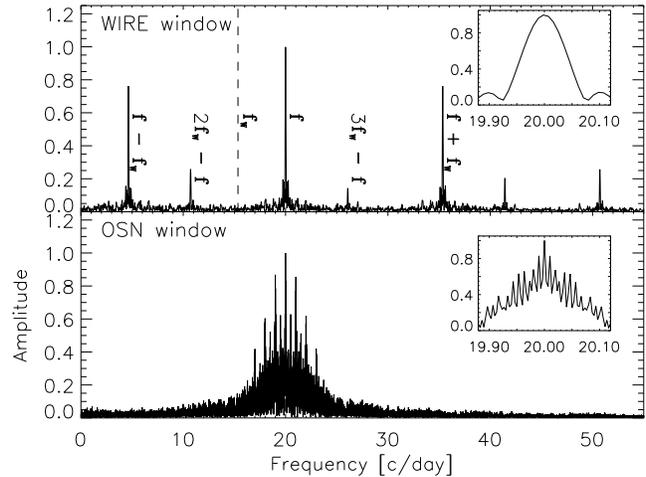}
   \caption{Spectral windows for the \wire\ (\topp\ \panel) and \osn\ data~sets (\bott\ \panel) 
            computed for a single frequency at $f = 20$ c/day. The insets
 show the details of the main peak.
              \label{fig:window}}
    \end{figure}

To illustrate the difference between 
the \wire\ and \osn\ data~sets in the frequency domain
we have calculated the spectral windows. To do this we used the same 
observation times as in the real data~sets and inserted an artificial 
sinusoidal signal at $f=20$ c/day. The resulting spectral windows for 
the \wire\ and \osn\ data~sets are shown in Fig.~\ref{fig:window} 
in the \topp\ and \bott\ \panels, respectively.

The complexity of the \osn\ spectral window is apparent,
with several alias peaks at 1.0, 0.5, and 0.01 c/d. 
The latter is seen in the inset in Fig.~\ref{fig:window} and arises from the 
large gaps in the \osn\ time series, \ie\ $f \simeq 1/T_{\rm obs} \simeq 0.01$ c/day, 
since the total observing time is $T_{\rm obs} = 122$\,days. 
Note that we decided not to use the last night from \osn\ in the analysis 
since it degrades the spectral window. The reason is the
long gap of 42 nights from night 15 to night 16.

As a result of the long gaps in the \osn\ time series,
the peaks in the amplitude spectrum at $f\pm n \times 1/T_{\rm obs}$ --
where $n$ is an integer -- are almost equally good solutions. 
However, if one can be sure about selecting the 
``right peak,'' the accuracy of the frequency is significantly 
better than in the \wire\ data~set. 
This may prove difficult since in the real data~set the spectrum
is affected by closely spaced frequencies and noise sources 
such as photon shot noise and non-white 
instrumental drift noise.

The \wire\ spectral window has a much more well-defined peak. 
During each \wire\ orbit the satellite switches between two targets in order to
minimize the effect of scattered light from the illuminated face of the Earth.
Thus, \epscep\ was observed with a duty cycle 
of $\simeq 40\%$ (\cf\ \bott\ \panel\ in Fig.~\ref{fig:lc}).
As a consequence, significant alias peaks are
seen at frequencies that are combinations of the frequency of the oscillation signal 
and the orbital frequency of \wire, $f_i = |f \pm n f_W|$, where 
$f$ is the genuine frequency, $n$ is an integer, and 
$f_W$ is the orbital frequency: $f_W=15.348\pm0.001$ c/day. The first set of
side lobes have amplitudes relative to the main peak of 76\%.

\subsection{Observed amplitude spectra}

The observed amplitude spectra are shown in Fig.~\ref{fig:amp}. 
In the two \topp\ \panels\ we have used the \wire\ data~set: the first
\panel\ is an overview and the second \panel\ shows the details 
of the region 12--28 c/day where 
the oscillations intrinsic to the star are found.
In the two \bott\ \panels\ we used 15 nights from \osn\ in the 
$y$ filter and $b-y$ colour, respectively.
The different properties of the time series are 
reflected in the amplitude spectra.

There are many frequencies present in \epscep\ and 
their location can be identified in the \wire\ spectrum. 
The periods range from 0.7--1.9 hours
and the amplitudes of highest peaks
are in the range 1--3 ppt\footnote{We note that $1$\,ppt $\simeq 1.086$\,mmag}, which is 
typical for low amplitude \dss\ stars.
The \osn\ amplitude spectra are more complicated to interpret,
with two main regions of excess power around 12--17 and 24--28 c/day.
This makes the extraction of closely spaced frequencies difficult
and is a well-known problem for single-site observations of \dss\ stars \citep{costa03}.
For example, as a result of the combination of the frequencies
$f_2$ to $f_4$, the highest peak in the \osn\ amplitude spectra 
is found at $\simeq 14$ c/day. Also, above 20 c/day the 
highest peak in the \osn\ $y$ amplitude spectrum is found around 
25.2 c/day due to the combinations of $f_1$ and the close pair
of frequencies $f_5$ and $f_9$.

In the following Section we describe how we have extracted
the individual frequencies from the light curves.


\section{Analysis of the \epscep\ light curves\label{sec:modes}}

\subsection{Using the superior \wire\ spectral window\label{sec:lcwire}}

Since the spectral window of the \wire\ data~set is less complicated than
for the \osn\ data, we used the \wire\ data to detect the significant
frequencies. Also, the S/N level is much higher in the \wire\ data:
in the cleaned amplitude spectrum the noise level is about 85 and 460 ppm in
the range 10--30 c/day for the \wire\ and $y$ \osn\ data, respectively.
In the \osn\ $v$ filter the noise level is 645 ppm but the amplitudes are
about 50\% higher than in $y$. The \wire\ data~set is useful for avoiding
the 1 c/day aliasing problem that hamper our single-site 
ground-based data~set and we can also detect additional frequencies with low amplitude.

We used the software package \period\ by \citet{lenz05}
for the extraction of the frequencies. After the extraction of the first
frequency, the detection of additional frequencies is based
on prewhitening or ``cleaning'' of the already detected frequencies.
However, the solution is
improved by a least-squares fit to the observations by a function of the
form $\Sigma_{i=1}^N A_i \sin (2 \pi [f_i~t + \phi_i]\,)$, 
thus each of the $N$ terms
is determined by frequency ($f_i$), phase ($\phi_i$) and amplitude ($A_i$).
We note that the \wire\ data~set is very homogeneous 
and we therefore did not apply point weights.

The \wire\ satellite observed \epscep\ during 40\% of its orbit,
but the last part of each orbit is affected by scattered light 
which systematically offsets the measured flux.
To minimize the effect of this we only used the part of the light curve which was  
unaffected by scattered light, \ie\ this data~set had a 30\% duty cycle.
From this data~set we extracted 25 frequencies with S/N above 4.
After subtracting these terms from the light curve we 
performed a decorrelation of the light curve with
the background level and orbital phase.
This allowed us to use the complete data~set and increase 
the duty cycle from 30\% to 40\%. 
This greatly improves the spectral window and the first set of 
side lobes decrease from 88\% to 76\% while the second set of 
side lobes decrease from 58\% to 25\% (\cf\ \topp\ \panel\ in Fig.~\ref{fig:window}).

\begin{figure*}
   \centering
\includegraphics[width=16.4cm]{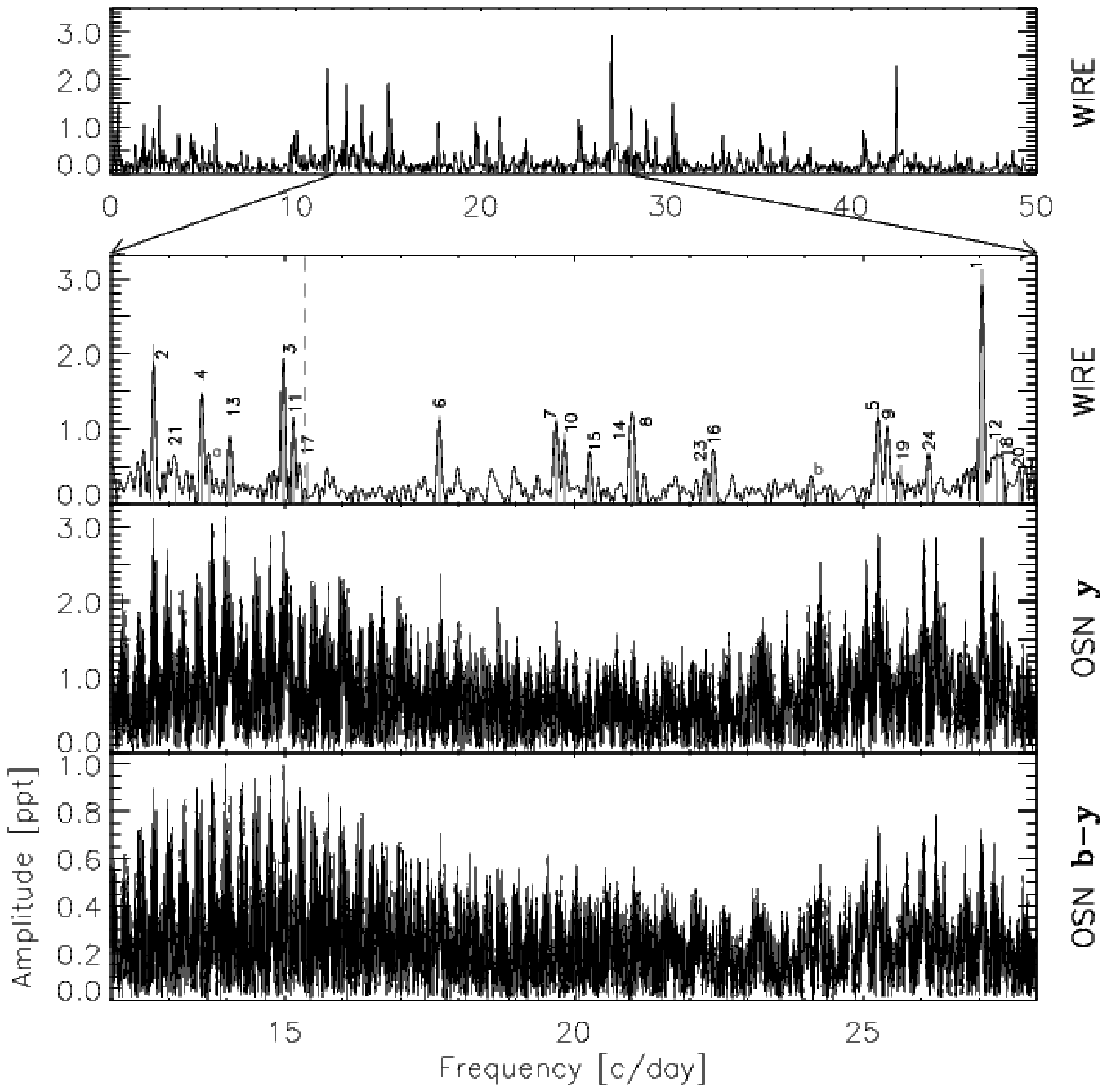}
   \caption{The two \topp\ \panels\ are the amplitude spectra 
of \epscep\ based on the \wire\ data~set. 
The frequencies extracted from \wire\ and the 
orbital frequency at $f_W = 15.348$ c/day (dashed line) have been marked.
The peaks below 12\,\cday\ and above 28\,\cday\ are alias peaks due to the orbit of the \wire\ satellite.
The two \bott\ \panels\ are the amplitude spectra of the \osn\ $y$ and $b-y$ data~sets. 
[{\em {\tt astro-ph}: quality of figure reduced.}]
              \label{fig:amp}}
    \end{figure*}

Using the \wire\ data~set with 40\% duty cycle we extracted 26 frequencies. 
The frequencies are marked in Fig.~\ref{fig:amp} and in Table~\ref{tab:freqwire} 
we list the frequency, amplitude, and phase of each frequency. 
Phases in Table~\ref{tab:freqwire} are given relative 
to the zero point in time, $t_0 =  2\,453\,175.5$. 
In the last column we give the S/N which is the ratio of the amplitude 
and the noise level estimated in the cleaned amplitude spectrum. 
We will estimate the uncertainties on the 
frequency, phase, and amplitude based on simulations in Sect.~\ref{sec:simul}.
The first 24 frequencies in Table~\ref{tab:freqwire} have S/N above 6, 
and are numbered according to their S/N. The remaining two frequencies are less 
certain are labeled $a$ and $b$.

We have searched for frequencies that are given as linear combinations 
of other terms within the frequency resolution, and for
the \wire\ data~set this is $\delta f = 3/2 T_{\rm obs} = 0.11$\,c/day \citep{loumos78}.
The three frequencies $f_{20}$, $f_{22}$, and $f_a$ are found to have low amplitude 
and found near these linear combinations:
$f_{20} = f_2 + f_3$, $f_{22}=f_{10}+f_{13}$, and 
$f_{a} = f_{12}-f_4 = f_{20} - f_{13} = f_{22}-f_{15}$.
The many combinations for $f_a$ indicate that it is probably not intrinsic to the star.
In addition, we find $f_9 = 2 \cdot f_2$ and $f_{18} = 2 \cdot f_{a}$, 
which may be chance alignments.


\begin{table}
\caption{Frequencies, amplitudes, phases, and S/N for 
26 individual frequencies extracted from the \wire\ data~set.
\label{tab:freqwire}}
\centering
\begin{tabular}{r|rrrr}
\hline\hline       
ID & $f_i$ [c/day] & $a_i$ [ppt] & $\phi_i$ & S/N \\
\hline

             $f_{1}$ &  27.053 &  3.12 &  0.994 & 41.2 \\ 
             $f_{2}$ &  12.734 &  2.13 &  0.329 & 27.6 \\ 
             $f_{3}$ &  14.976 &  1.80 &  0.524 & 23.4 \\ 
             $f_{4}$ &  13.568 &  1.36 &  0.617 & 17.7 \\ 
             $f_{5}$ &  25.262 &  1.25 &  0.096 & 16.1 \\ 
             $f_{6}$ &  17.674 &  1.18 &  0.238 & 15.4 \\ 
             $f_{7}$ &  19.689 &  1.13 &  0.858 & 14.6 \\ 
             $f_{8}$ &  21.041 &  1.03 &  0.068 & 13.3 \\ 
             $f_{9}$ &  25.409 &  0.99 &  0.541 & 12.8 \\ 
            $f_{10}$ &  19.842 &  0.95 &  0.756 & 12.3 \\ 
            $f_{11}$ &  15.157 &  0.90 &  0.939 & 11.8 \\ 
            $f_{12}$ &  27.298 &  0.86 &  0.618 & 11.3 \\ 
            $f_{13}$ &  14.050 &  0.86 &  0.573 & 11.2 \\ 
            $f_{14}$ &  20.980 &  0.84 &  0.771 & 10.9 \\ 
            $f_{15}$ &  20.255 &  0.70 &  0.781 &  9.1 \\ 
            $f_{16}$ &  22.415 &  0.64 &  0.711 &  8.1 \\ 
            $f_{17}$ &  15.392 &  0.55 &  0.489 &  7.2 \\ 
            $f_{18}$ &  27.416 &  0.52 &  0.872 &  6.9 \\ 
            $f_{19}$ &  25.663 &  0.53 &  0.013 &  6.8 \\ 
  $f_2+f_3 = f_{20}$ &  27.702 &  0.49 &  0.538 &  6.5 \\ 
            $f_{21}$ &  13.118 &  0.48 &  0.131 &  6.3 \\ 
$f_{10}+f_{13}=f_{22}$& 33.957 &  0.44 &  0.793 &  6.3 \\ 
            $f_{23}$ &  22.277 &  0.48 &  0.101 &  6.1 \\ 
            $f_{24}$ &  26.119 &  0.47 &  0.202 &  6.1 \\ 
           $f_{a}^1$ &  13.697 &  0.39 &  0.486 &  5.0 \\ 
             $f_{b}$ &  24.102 &  0.35 &  0.865 &  4.5 \\ 


\hline                    
\hline                  
\noalign{\smallskip}
\multicolumn{5}{l}{$^1$ Note that $f_a=f_{12}-f_4=f_{20}-f_{13}=f_{22}-f_{15}$}
\end{tabular}
\end{table}

\subsection{Analysis of the \osn\ light curves \label{sec:lcosn}}

The \osn\ data~set was collected 2.9 years prior to the \wire\ data~set.
While some \dss\ stars are known to have variable amplitudes 
we expect
that the frequencies remain constant over such a short time scale.
However, \cite{breger06} found that the amplitude variation
seen in FG~Vir during several observing seasons can be explained 
by closely spaced frequencies.
With this in mind we searched for frequencies using \period\ while
using the frequencies extracted from \wire\ as a guide. 
We recovered the frequencies $f_1$ to $f_9$, but in some cases we had to apply 
an offset of the apparently highest peak by exactly $\pm1$\,c/day.
We found evidence for $f_{11}, f_{12}$, and $f_{19}$ but these frequencies have
low amplitude and the systematic offsets due to close neighbours
and their aliases become significant.

Due to the long gaps in the \osn\ data~sets there are aliases in the
spectral window separated by 
$f_{\rm obs} \simeq 1/T_{\rm obs} = 0.009$\, c/day. These aliases have
almost equal amplitude but we can pick the right peak in the amplitude
spectrum using the approximate frequency from the \wire\ data~set
as we will demonstrate in Sec.~\ref{sec:ambiguity}.
From the simulations in Sec.~\ref{sec:simul} we find
the uncertainty on the \wire\ frequencies
to be $0.001-0.003$\,c/day for the frequencies $f_1$ to $f_6$ and up to $0.004$\,c/day
for the frequencies $f_7$ to $f_9$. This means that the shift from one alias
to the next in the \osn\ amplitude spectrum is at least 
at the 3-sigma level for $f_1$ to $f_6$ and about 2-sigma for $f_7$ to $f_9$. 

\begin{figure}
   \centering
\includegraphics[width=8.8cm]{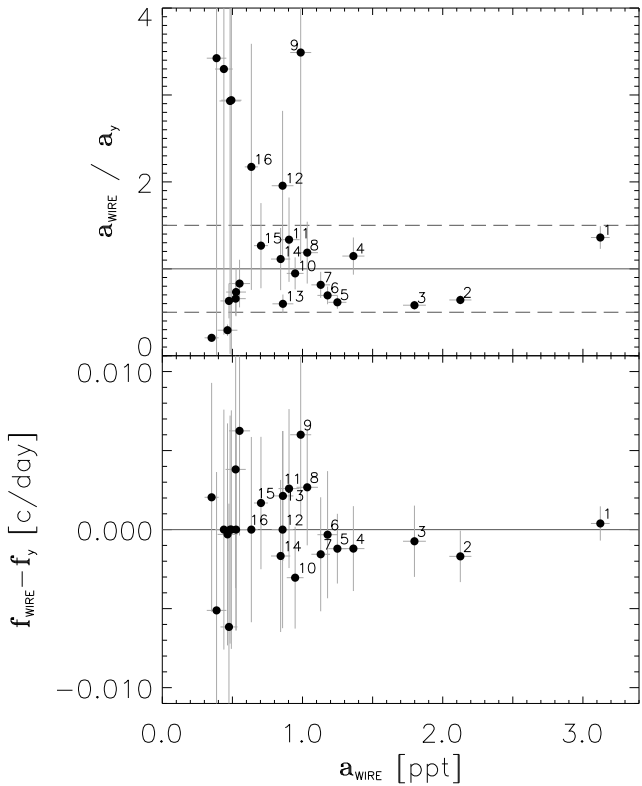}
      \caption{We fitted all 26 frequencies found 
from the \wire\ data~set to the \osn\ $y$ data~set. 
The \panels\ show the ratio of the input and output amplitudes (\topp)
and the difference between input and output frequencies.
              \label{fig:wireosn}}
    \end{figure}

Another method of cleaning the \osn\ data~set 
is to assume that all frequencies with S/N above 6 
found in \wire\ can be fitted to the \osn\ data~set.
In Fig.~\ref{fig:wireosn} we compare the frequencies and 
amplitudes found in the \wire\ and \osn~$y$ data sets. 
The \topp\ \panel\ in Fig.~\ref{fig:wireosn} shows 
the ratio of amplitudes $a_{\rm WIRE}/a_{\rm y}$. 
The \bott\ \panel\ shows the difference between the frequencies \vs\ the \wire\ amplitude. 
The individual error bars are found from simulations.
We find that the uncertainties are very large for the frequencies with low amplitude,
but all frequencies agree within the uncertainties. That fact that several
frequencies are found to have amplitudes that are different by more than 50\% 
(marked by horizontal dashed lines in the \topp\ \panel\ in Fig.~\ref{fig:wireosn})
indicates that it is not sensible to fit all these frequencies to the
\osn\ data, although they are present in the \wire\ data~set.
The fact that we cannot extract all the frequencies found in the \wire\ data~set
will systematically affect the parameters of the frequencies we extract
from the \osn\ data~sets. 
We have assessed this by doing simulations (see Sect.~\ref{sec:simul}).

In Table~\ref{tab:freqcomb} we summarize the frequency, amplitude, and phase of $f_1$ to $f_8$,
which are clearly identified in the \osn\ data~set.
The frequencies are the weighted mean values of the individual
fits to $uvby$. 
The quoted error is the weighted mean error and
is based on simulations done in Sect.~\ref{sec:simul}. 
We give amplitudes in each of the four \str\ filters and
the colour light curve $b-y$. The phases of $y$ and $b-y$ are given
relative to the zero point in time, $t_0 =  2\,452\,130.0$.
Three of the frequencies, $f_5, f_7$, and $f_8$ have some closely spaced
frequencies seen in the \wire\ data~set, namely $f_9, f_{10}$, and $f_{14}$.
It is very likely that the parameters listed 
for these frequencies in Table~\ref{tab:freqcomb} are
systematically affected by this. 
We also note that $f_7\simeq f_6+2.0$\,c/day
and this will also affect the amplitude and phase.

%

   \begin{figure}
   \centering
\hskip -0.41cm \includegraphics[width=9.2cm]{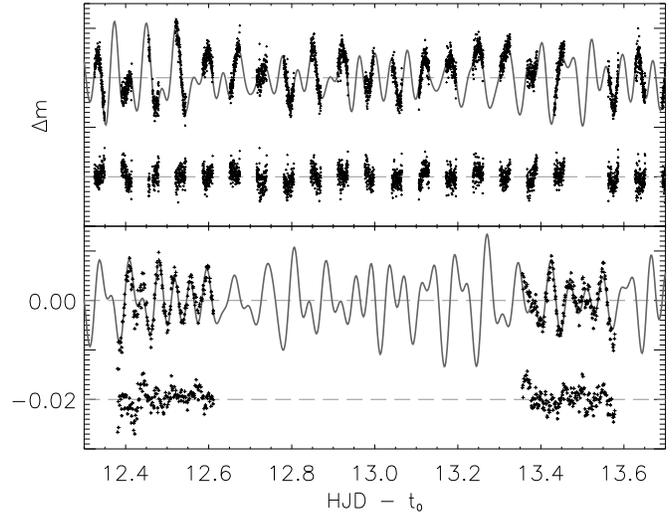}
   \caption{Part of the light curves from \wire\ (\topp\ \panel)
and \osn\ $y$ including the fitted light curves in grey colour. 
The residuals are also shown but offset by $-0.02$~mag. 
As in Fig.~\ref{fig:lc} the zero-points $t_0$ are different.
              \label{fig:fitlc}}
    \end{figure}


\section{Accuracy of the extracted frequencies\label{sec:freq}}

\subsection{Theoretical estimates\label{sec:theory}}

The formal uncertainty
on the frequency determined from a light curve is determined by 
the duration of the observing run, 
the number of data points, and S/N ratio, 
\ie\ the ratio of the amplitude to the noise in the light curve.
The standard estimate 
of the uncertainty of the frequency is based on the
least-squares covariance matrix or the Rayleigh
resolution criterion.
However, \cite{schwarzenberg91}
demonstrated that both estimates are statistically incorrect.
On one hand the least-squares covariance matrix does not account
for correlation of residuals in the fit. Neglecting
this may cause a large under-estimation of the uncertainty.
On the other hand the Rayleigh resolution criterion
is insensitive to the S/N and therefore
does not reflect the quality of the observations.

For an {\em ideal} light curve with only white noise \cite{montgomery}
derived the uncertainty on the frequency, amplitude and 
phase (in radians) as

  \begin{equation}
  \sigma(f_i) = 
   {{\sqrt{6}} \over {\pi}} \cdot {1 \over {N^{1/2}}} \cdot {1 \over {T_{\rm obs}}} \cdot {\sigma \over {a_i}},
   \label{eq:sigmaf}
  \end{equation}
  \begin{equation}
  \sigma(a_i) = {{\sqrt{2 \over {N}} \cdot \sigma}} \hskip 0.3cm {\rm and}
   \label{eq:sigmaa}
  \end{equation}
  \begin{equation}
  \sigma(\phi_i) = {{\sqrt{2 \over {N}} \cdot {{\sigma} \over {a_i}}}},
   \label{eq:sigmap}
  \end{equation}
where $\sigma$ is the \rms\ uncertainty per data point,
$a_i$ is the amplitude, $N$ is the number of data points, 
and $T_{\rm obs}$ is the observational time baseline. 
These uncertainties are strictly lower limits in the case of 
uncorrelated white noise.
Using simple assumptions for correlated noise, 
\cite{montgomery} found that the error estimates using the above 
equations are too optimistic by up to a factor of five.
Following the suggestion by \cite{montgomery}
we calculated the {\em correlation length} $D$ from the 
autocorrelation of the cleaned amplitude spectra (after subtracting the mean) 
and located the intersection with zero. 
From the \wire\ and \osn\ data~sets we find $D_{\rm WIRE}\simeq10.5$, 
$D_{{\rm OSN;}y}=9.0$, and $D_{{\rm OSN;}b-y}=10.5$,
respectively, and we multiplied the error 
estimates from Eq.~\ref{eq:sigmaf}--\ref{eq:sigmap} 
by the square root of the corresponding correlation length, $D$.

%
   \begin{figure}
   \centering
\includegraphics[width=8.8cm]{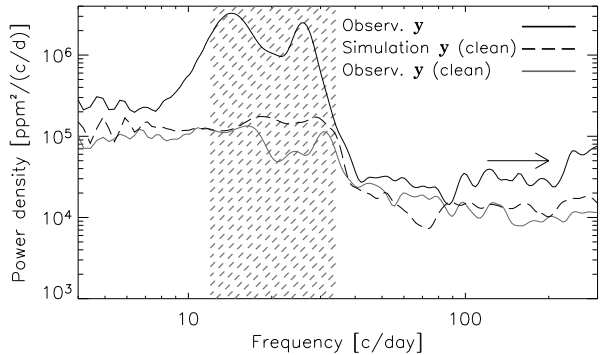}
   \caption{Smoothed power density spectrum of the \osn\ $y$ data~set before and after cleaning (solid lines).
The dashed line is for a simulation after cleaning. 
The location of the oscillations is marked by the hatched region.
              \label{fig:density}}
    \end{figure}

\subsection{Power density spectra\label{sec:pd}}

For the simulations done in Sect.~\ref{sec:simul} 
we have found that it is important that 
the noise sources in the simulations
mimic the observed data as closely as possible 
and we will investigate them in detail here.

In Fig.~\ref{fig:fitlc} we show part of the 
light curve of \epscep\ from \osn\,$y$ and \wire.
The grey curves show the fit to each of the complete light curves.
The residuals are also shown, offset by $-0.02$\,mag.
It is seen that there are significant systematic trends in the
residuals which may be due to a combination of 
unresolved frequencies and instrumental drift.

The significance of this is seen more clearly in the 
frequency domain, and in Fig.~\ref{fig:density} we show
a smoothed version of the power density (PD) spectra of the \osn\,$y$ data~set
before and after the cleaning process. 
The region of excess power due to the oscillations is indicated 
by the hatched region (12--35 c/day). 
The PD is seen to increase by about an order of magnitude from the theoretical 
white noise level ($>80$\,c/day). 
The slight increase in noise level
above 200\,c/day (marked by the arrow) is because the 
telescope at \osn\ switched between the \epscep\ and 
the comparison stars every 2--4 minutes.

\begin{figure}
   \centering
\includegraphics[width=8.8cm]{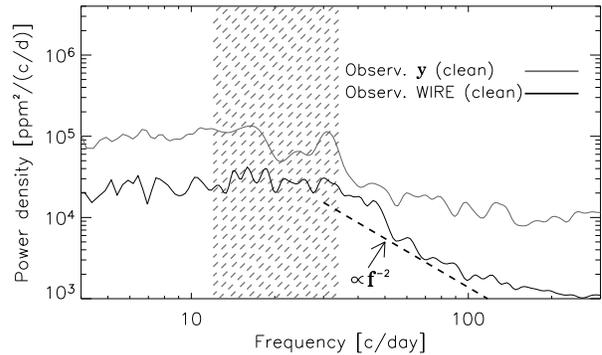}
   \caption{Smoothed power density spectra of the cleaned \osn\ and \wire\ data~sets.
The dashed line indicates the increase in noise towards low frequencies.
              \label{fig:density2}}
    \end{figure}

A virtue of the PD spectrum is that we can compare 
the frequency dependence of the noise in different
data~sets even though the temporal coverage,
time sampling, and number of data points are quite different.
In Fig.~\ref{fig:density2} we compare the 
PD spectra of the cleaned \osn\,$y$ (grey) 
and \wire\ spectra (black solid line).
The curves are similar in shape but there are fundamental differences. 
At the high frequency end the noise in \osn\,$y$ is higher 
by an order of magnitude and this is because the \wire\ data~set has
$\sim10$ times more data points than \osn\ and slightly lower
uncertainty on each data point. 
At low frequencies ($f<40$\,c/day) the PD is still 
smaller for \wire\ at all frequencies which means that there are 
additional noise sources present in the \osn\ data~set.
If the noise were intrinsic to the star
the PD in \wire\ and \osn\ would be identical at
low frequencies.
However, a large number of frequencies is present in \epscep\ and 
a perfect cleaning of the \osn\ data~set is not possible. 
This probably explains the higher PD at $f<40$\,c/day 
compared to \wire\ by a factor of $\simeq3$.

%
\begin{table*}
\caption{Frequencies and amplitudes in the \str\ filters for frequencies extracted 
from the \osn\ data~set. The frequencies are the weighted mean of the fit 
to each of the $uvby$ light curves.
Amplitudes are given for each filter in parts per thousand (ppt). 
Phases fitted to $y$ and $b-y$ are also given and indicated in units of the period.
The three frequencies labeled $cl$ have a close neighbouring frequency with similar amplitude and
the parameters are likely to be affected by this.
\label{tab:freqcomb}}           
\centering
\begin{tabular}{l|r|rrrrrr|rr}
\hline
\hline       
ID       & $f_{{\rm comb}}$ [c/d]   & $a_W$ & $a_u$ & $a_v$ &  $a_b$ &  $a_y$ & $a_{b-y}$ & $\phi_y$ & $\phi_{b-y}$ \\
\hline
 $f_{  1}$ &   $   27.0522\pm  0.0003$ &   3.12&   2.76 & 3.12&   2.61 &  2.13 &  0.496  &  0.570 &   0.584 \\
 $f_{  2}$ &   $   12.7357\pm  0.0002$ &   2.13&   4.15 & 5.15&   3.78 &  3.19 &  0.713  &  0.046 &   0.130 \\
 $f_{  3}$ &   $   14.9766\pm  0.0002$ &   1.80&   3.78 & 4.61&   3.79 &  2.82 &  1.085  &  0.447 &   0.470 \\
 $f_{  4}$ &   $   13.5688\pm  0.0009$ &   1.36&   1.80 & 2.12&   1.64 &  1.17 &  0.459  &  0.373 &   0.462 \\
 $f^{cl}_{  5}$ & $   25.2635\pm  0.0004$ &   1.25&   2.88 & 3.33&   2.72 &  2.13 &  0.654  &  0.139 &   0.159 \\
 $f_{  6}$ &   $   17.6745\pm  0.0004$ &   1.18&   1.46 & 2.19&   2.11 &  1.81 &  0.514  &  0.955 &   0.705 \\
 $f^{cl}_{  7}$ & $   19.6904\pm  0.0004$ &   1.13&   1.27 & 2.00&   1.89 &  1.34 &  0.639  &  0.315 &   0.106 \\
 $f^{cl}_{  8}$ & $   21.0380\pm  0.0004$ &   1.03&   1.99 & 1.92&   1.43 &  1.04 &  0.387  &  0.028 &   0.889 \\

\end{tabular}
\end{table*}

   \begin{figure*}
   \centering
\includegraphics[width=8.8cm]{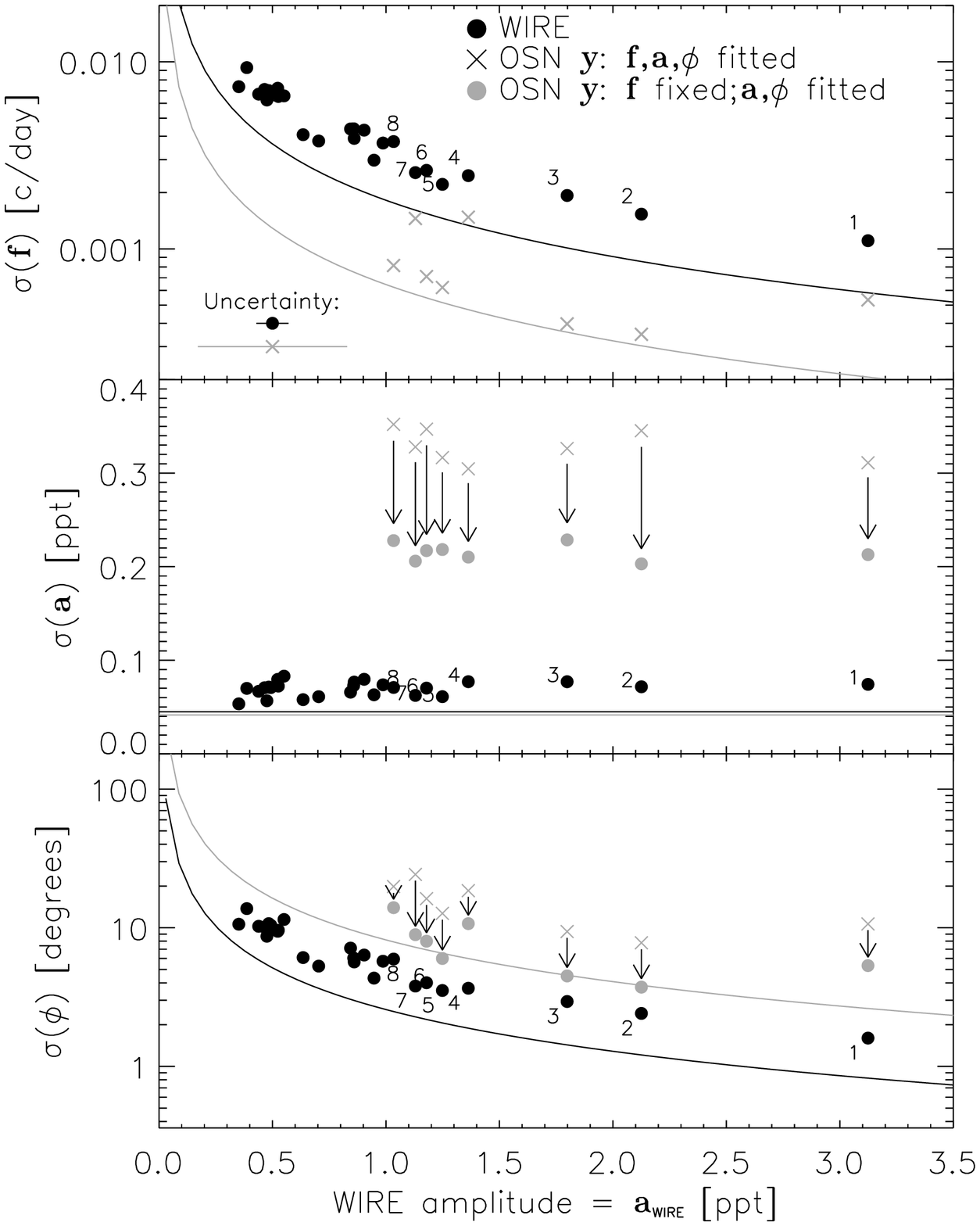}
\includegraphics[width=8.8cm]{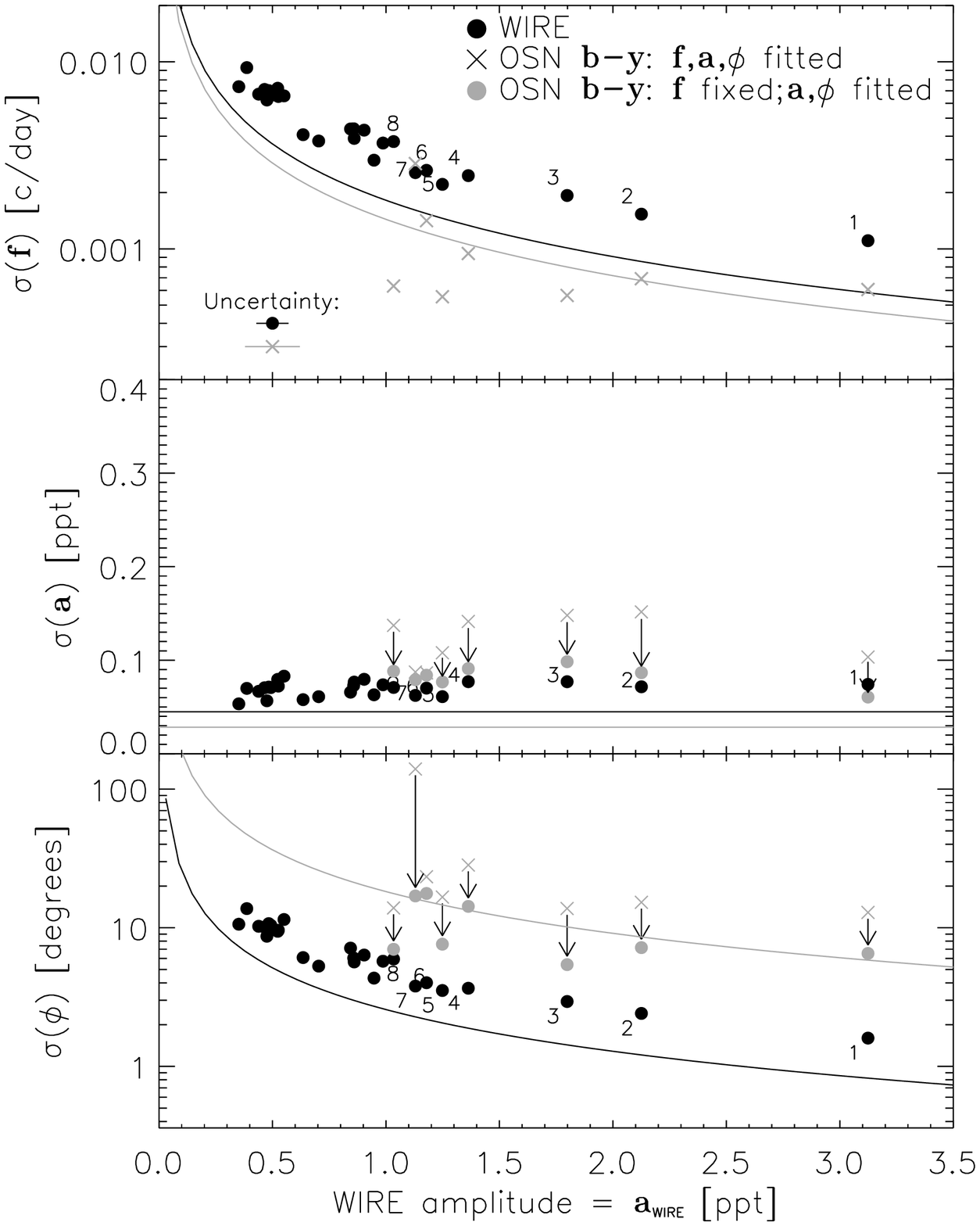}
      \caption{Uncertainties on frequency, amplitude, and phase as
found from simulations of \wire\ (black) and \osn\ (grey) data~sets.
The \wire\ results are shown in both \panels\ and the \osn\ results for $y$ and
$b-y$ are shown in the \lee\ and \rii\ \panels, respectively.
When fixing the frequencies in the \osn\ data~set the uncertainties on the 
amplitudes and phases decrease as indicated by arrows.
The solid lines are the theoretical
predictions for the uncertainty from Eqs.~\ref{eq:sigmaf}--\ref{eq:sigmap}.
              \label{fig:uncertain}}
    \end{figure*}

The increase towards lower frequencies in the \wire\ data~set is 
likely due to a combination of instrumental drift and a 
number of undetected frequencies.
This possibility was also discussed by \cite{breger05} in their 
analysis of the residuals after cleaning about 80 frequencies 
in the \dss\ star FG~Vir.
We motioned in Sect.~\ref{sec:previous}
that \citet{kennelly99} found several frequencies of high degree based on
LPV studies. These frequencies will have negligible amplitude in 
photometry due to geometrical cancellation effects, but they 
may produce part of the increase in the noise that we observe.

\subsection{Simulations of the \wire\ and \osn\ time series\label{sec:simul}}

To confirm the theoretical error estimates in Sect.~\ref{sec:theory}
and to better understand the
different properties of the \osn\ and \wire\ data~sets 
we computed a large set of simulations.
The \osn\ data~set has long gaps of up to a month and thus the 
frequency analysis is hampered by a complicated spectral window.
The interaction between frequencies can only be estimated by doing
a large number of realistic simulations.

To mimic the large increase towards low frequencies discussed in Sect.~\ref{sec:pd},
we added a number of frequencies with low amplitude in a wide frequency range. 
The simulations of the \osn\ $y$ filter and $b-y$ data~set were done 
by including the 26 frequencies detected in \wire\ but using the frequencies, 
amplitudes, and phases fitted to the observed data.
We then added 110 frequencies with random (low) amplitudes from 0.0 to 0.4\,ppt 
and frequencies in the range 1--35\,c/day for the $y$ filter. 
For the $b-y$ light curve we added 350 frequencies with amplitudes 0.0--0.1\,ppt in the same frequency range.
Finally, we added a white noise component with \rms\ of $1.85$ and $1.60$\,mmag
for $y$ and $b-y$, respectively.
In the simulations of the \wire\ data~set we add 75 frequencies with
random frequencies in the range 10--35\,c/day, 
random amplitudes in the range 0.0--0.3 ppt, random phases, 
and a white noise component with \rms\ $1.7$\,mmag. 
An example of the cleaned PD spectrum of a 
simulation of an \osn\ $y$ time series is  
shown with a dashed line in the \lee\ \panel\ in Fig.~\ref{fig:density}.

These {\em ad hoc} simulations roughly reproduce the increase
in noise towards low frequencies while still no significant frequencies
are present above the noise. 
We made 800 simulations of the \osn\,$y$, $b-y$, and \wire\ light curves. 
We extracted frequencies from all simulations using an automated cleaning 
program using the procedure described in Sect.~\ref{sec:lcwire}. 
For the \osn\ data~set we also fitted the light curves when assuming
the known frequencies and only fitting amplitudes and phases. 
We found the uncertainty by calculating 
the \rms\ scatter in frequency, amplitude, and phase extracted from
the 800 simulations.

   \begin{figure*}
   \centering
\includegraphics[width=17.4cm]{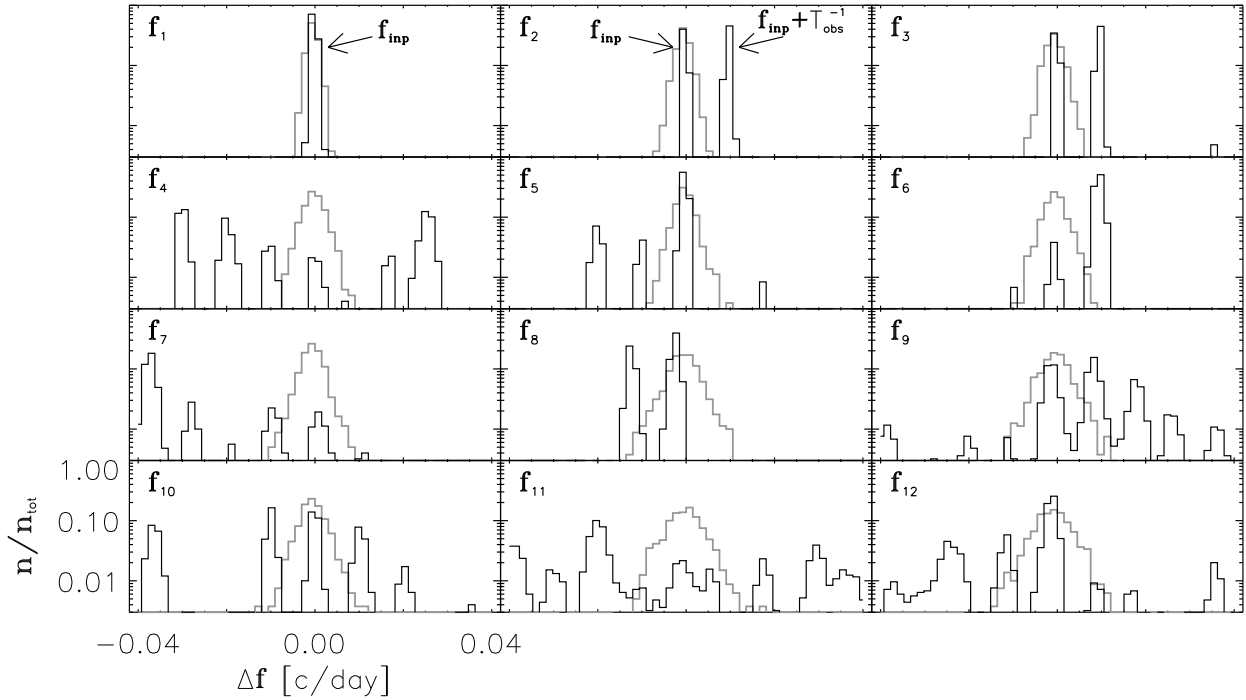}
      \caption{Histogram of frequencies recovered in simulations of the
\osn\ (black) and \wire\ (grey) data~sets for $f_1$ to $f_{12}$.
In the \panel\ for $f_2$ two groups of 
solutions in \osn\ separated by $1/T_{\rm obs}=0.009$\,c/day are marked.
It is seen that the \wire\ data~set can be used to pick the right group, 
and that within this group the \osn\ data~set has significantly lower uncertainty.
              \label{fig:histofreq}}
    \end{figure*}

In Fig.~\ref{fig:uncertain} we compare the uncertainties
on frequency, amplitude, and phases as found from 
Eq.~\ref{eq:sigmaf}--\ref{eq:sigmap} (solid lines) and the 
simulations (points).
The \lee\ \panel\ shows the uncertainties on the 
frequency, amplitude, and phase
for \osn\,$y$ and the \rii\ \panel\ is for \osn\,$b-y$, while
the results for \wire\ are shown with black points in both \panels. 
The uncertainties of the simulations of the \osn\ data~sets are shown 
when both frequency, amplitude and phase are fitted (grey $\times$ symbols)
and the systematically lower uncertainties when only amplitude 
and phase are fitted (grey circle symbols). 
The improvement is indicated by a vertical arrow for each frequency.
It can be seen that the frequencies are more 
accurately determined when using the \osn\,$y$ data~set. 
This is in agreement with Eq.~\ref{eq:sigmaf} where
the time baseline enters linearly while the number of data points
enters as the square root; for the two data~sets we approximately 
have the ratios $N_{\rm WIRE}/N_{\rm OSN} \simeq T_{\rm OSN}/T_{\rm WIRE}\simeq10$,
while the point-to-point noise, $\sigma$, is very similar.
For the frequencies $f_1$ to $f_8$ the errors on frequency are 
about $0.001-0.003$\,c/day in \wire\ and 
$0.0005-0.0010$\,c/day in \osn.
We should note that the \osn\ results rely on 
the important fact that we can select the right 
alias peak in the \osn\ data by using the {\em approximate} frequency 
found with \wire; this is discussed in detail in Sect.~\ref{sec:ambiguity}. 
The uncertainties on amplitude and phase
are independent of the time baseline and therefore they are more 
accurately determined from 
the \wire\ data~set since $N_{\rm WIRE}/N_{\rm OSN}\simeq10$.

\subsection{Resolving the ambiguity of \osn\ alias peaks\label{sec:ambiguity}}

Due to long gaps in the \osn\ data~set the spectral window has aliases 
of similar amplitude separated by $\Delta f\simeq0.009$\,c/day. 
If we only had the \osn\ data~set, each frequency found in
the observed data~set can be offset by $n \times 0.009$
c/day for any $n=\pm1,\pm2$. However, using the frequencies 
found from \wire\ we may choose the right alias {\em sub-peak} 
in the \osn\ amplitude spectrum.

The results from our simulations in 
Fig.~\ref{fig:histofreq} illustrate that this is indeed possible. 
Each \panel\ shows two histograms of the difference between 
the input frequency and the {\em extracted} frequency
for $f_1$ to $f_{12}$:
the grey histogram is for the simulations of the \wire\ data 
and black is used for \osn\,$y$. 
Several "groups" of frequencies separated by
0.009 c/day are seen for the \osn\ simulations while a single but
broader peak is seen for the distribution of extracted \wire\ frequencies.

It can be seen that the uncertainty in the \wire\ data~set is sufficiently small
that we can select the right peak in the \osn\ data~set, at least for the
dominant frequencies. 
We note that the uncertainties in the \osn\ simulations shown in Fig.~\ref{fig:uncertain} 
are based on the internal \rms\ scatter within one "group" of extracted frequencies.

\section{Comparison of observations and models}

%
\subsection{Amplitude ratios and phase differences\label{sec:phase}}

\begin{figure}
   \centering
\includegraphics[width=8.8cm]{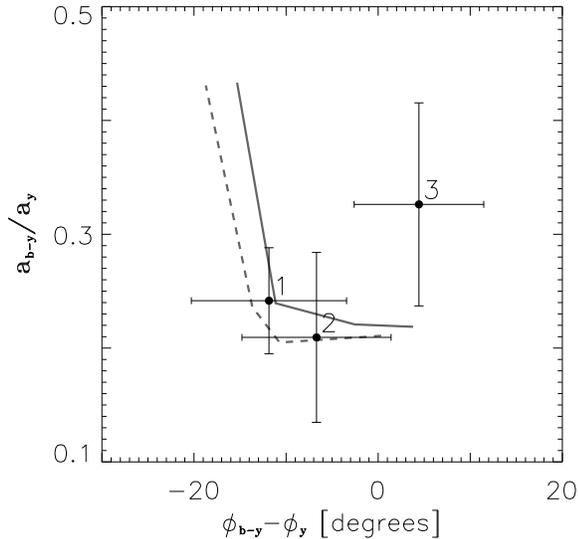} 
      \caption{Amplitude ratio and phase difference for the frequencies $f_1$ to $f_3$ 
measured in the \osn\ $y$ and $b-y$ light curves.   
The solid and dashed lines correspond to the fundamental mode 
and third overtone for a theoretical model with mass 
$M/M_\odot=1.65$.
              \label{fig:phaselag2}}
    \end{figure}

By measuring the parameters of frequencies in different filters we can infer
the spherical degree, $l$, of the associated spherical harmonic.
Each frequency can in principle be identified
by measuring the phase difference and amplitude ratio 
in two filters as shown by \citet{garrido90} and \citet{moya04}. 
In Fig.~\ref{fig:phaselag2} we show the
the amplitude ratio \vs\ the phase differences in the $y$ filter and $b-y$ colour
for the three frequencies $f_1$ to $f_3$.
The uncertainties are based on the simulations described in Sect.~\ref{sec:simul}.
The solid and dashed lines  are results   
for a model with mass $M/M_\odot=1.65$ and \teff\,$=7720$~K. 
We used an over-shooting parameter of $d_{\rm ov}=0.2$ but
did not include the effects of rotation.
We note that in Sec.~\ref{sec:param} we inferred a slightly cooler temperature,
\ie\ \teff\, $=7340\pm150$ and an evolutionary mass of $M/M_\odot=1.75\pm0.20$.
In the model shown in Fig.~\ref{fig:phaselag2} the solid 
line is for $Q=0.033$ (fundamental radial mode, $n=1$) and 
the dashed line for $Q=0.017$ (third overtone, $n=4$) for
a mixing length parameter $\alpha = 0.5$. 
Amplitude ratios and phase differences were calculated for $l=0,1,$ and $2$:
the highest amplitude ratio is for $l=2$ at $a_{b-y}/a_y>0.4$ and becomes
progressively lower for decreasing $l$.
The frequencies $f_1$ and $f_2$ are compatible with $l=1$ or $2$.

In general, the amplitude ratio and phase diagrams depend on the
assumed mixing length parameter \citep{moya04,dasz03}.
Although the model used here does not describe \epscep\ in detail, we can see that the
observational uncertainties on the amplitude ratios and phases are too
large to distinguish between the radial degree and overtone.

%
\subsection{Search for a pattern in the amplitude spectrum}

Regular frequency spacings similar to the large separation
seen in solar-like stars have also been reported
for some \dss\ stars \citep{handler00}.
Such a measurement would enable us to compare with theoretical
predictions. We used the 24 most significant frequencies seen in \epscep, namely
those with S/N above 6 to look for significant spacings using
autocorrelation and histograms of frequency differences for 
different bin sizes.

\begin{figure}
   \centering
\includegraphics[width=8.8cm]{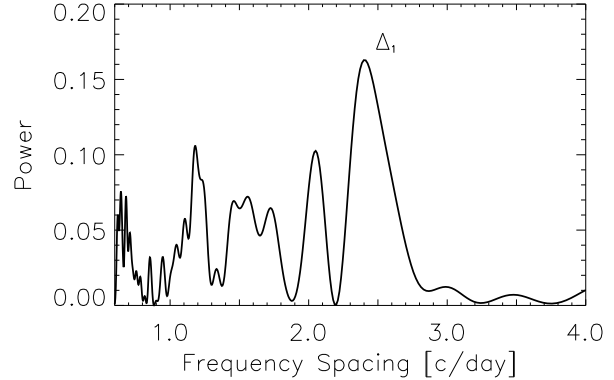}
      \caption{The power spectrum used to search for frequency spacings in \epscep. 
The 24 most significant frequencies found from the \wire\ data~set were used.
              \label{fig:spacing}}
    \end{figure}

In Fig.~\ref{fig:spacing} we plot the power \vs\ the spacing frequency.
We find a peak at $\Delta_1 = 2.41\pm0.02$ c/day.
We did a series of simulations to see if the $\Delta_1$ spacing 
is indeed significant. 
Each simulation consist of 24 frequencies 
randomly distributed in the range 12--28 c/day.
In many cases we found peaks in the amplitude spectrum with the same
approximate location and amplitude as $\Delta_1$.
Therefore we are extremely cautious about associating this with 
\eg\ half of the large separation. 
The low number of observed frequencies 
as well as the narrow frequency range that is covered, 
are insufficient to 
consider $\Delta_1$ as being statistically significant.

We also used another technique to look
for repetitive spacings among the frequencies 
by calculating frequency splitting histograms.
Several frequency binning widths were used in order to search for recurring peaks.
The only peak prevailing for all binning values is around 1.3\,c/day 
which roughly corresponds to the peak at 1.2\,c/day seen in Fig.~\ref{fig:spacing}.
This spacing can be interpreted as the rotational splitting or perhaps the small separation. 
In the region of large differences, when varying
the binning, two peaks are observed in the range of 4.8--5.0 and 5.7--5.9 c/day.
We are cautious about these large ``candidate'' splitting values, 
since we only observe frequencies in a narrow frequency range.
The frequencies cover 12.7--34.0\,c/day, so any spacing above half this range should
cannot be considered: $\Delta f_{\rm lim}=10.7$\,\cday. 
Even at 5\,c/day we would only be able to detect 
an even spacing of four consecutive frequencies.

We computed a number of pulsation models within 
the photometric error box for \epscep\ (\cf\ Fig.~\ref{fig:hr}). 
We selected those models consistent with $\Delta_1$ being 
half the value of the mean large separation, 
\ie\ $\Delta \nu \simeq 4.8$ c/day.
The parameters of these models are given in Table~\ref{tab:models}. 
For each model we give the mass, radius, and luminosity in solar units, 
the effective temperature,
the fractional hydrogen content in the central region, 
the large spacing of the frequencies, 
and the mean stellar density in solar units. 
We find that the observed peak in the histogram at $\simeq1.3$\,c/day 
is unlikely to be due to rotational splitting, since this peak 
is only seen in histograms of theoretical models for $m=0$ modes.
Thus, if the reality of this spacing can be established from a more
ambitious campaign, this splitting can only correspond to the small separation.

%
\section{Discussion and outlook\label{sec:concl}}

We have analysed observations of the \dss\ star \epscep\ from
the \wire\ satellite and ground-based multi-colour $uvby$ photo\-metry from \osn.
In some respects the \wire\ data~set is superior 
to \osn\ because the S/N is a factor 5 higher and 
the spectral window is much more simple. Thus, we used the \wire\ data~set
to extract the location of the oscillation frequencies.
Due to the long gaps in the \osn\ light curve, the resulting 
spectral window is complicated. In addition to aliases
at 1.0 and 0.5 c/day, there are also ``fine structure'' sub-peaks with a 
spacing of 0.009 c/day (roughly $1/T_{\rm obs}$). 
We performed a large set of realistic simulations to demonstrate that
we can select the right sub-peak in the \osn\ amplitude 
spectrum, and in turn determine the frequencies of the dominant
frequencies with accuracies of just $\simeq0.0005$\,c/day. 
The frequencies found with each 
of the four \str\ filters agree and we have computed 
the weighted mean frequency, 
thus improving in accuracy to $\simeq0.0003$ c/day for the main frequencies
(\cf\ Tab.~\ref{tab:freqcomb}).

Accurate frequencies are only of interest if they can
be compared with theoretical models, and this requires
that the degree of the mode can be determined.
This can be done from the amplitude ratio and phase difference
in different filters. 
We used the amplitudes and phases from 
the \osn\,$y$ filter and $b-y$ colour light curve,
but our comparison with a theoretical model
clearly indicates that the observed accuracies of phases 
and amplitude ratios are insufficient to perform 
a mode identification.
While accurate frequencies can
be obtained very effectively by extending the time baseline 
of the observations, 
accurate phases and amplitudes require higher S/N.
The uncertainties we estimate from simulations tell us that 
around 25\,000 data points with $\simeq2$\,mmag point-to-point
uncertainty are required, but we only have 10\% of this available 
from \osn.

\begin{table}
\caption{Parameters of selected theoretical models for \epscep.
The mass, radius, and luminosity are given in solar units, 
\teff\ is the surface temperature, $X_c$ is the central
hydrogen fraction, $\Delta \nu$ is the large separation,
and $\rho$ is the stellar mean density in solar units.
\label{tab:models}}      
\centering
\begin{tabular}{c|cccccc}
   \hline\hline       
   $M/M_\odot$ & $R/R_\odot$ & $L/L_\odot$ & \teff~[K] & $X_c$ & $\Delta \nu_{l=0}$ [c\,d$^{-1}$] & $\rho/\rho_\odot$ \\
   \hline

   1.55 & 1.76 &  6.70 & 7000 & 0.48 & 4.99 & 0.283 \\
   1.60 & 1.77 &  7.50 & 7190 & 0.50 & 4.94 & 0.287 \\
   1.65 & 1.79 &  8.46 & 7370 & 0.51 & 4.95 & 0.288 \\
   1.70 & 1.82 &  9.57 & 7530 & 0.51 & 4.89 & 0.281 \\
   1.75 & 1.83 & 10.70 & 7720 & 0.52 & 4.90 & 0.284 \\
   1.80 & 1.84 & 11.89 & 7910 & 0.53 & 4.92 & 0.288 \\
   \hline \hline                  
\end{tabular}
\end{table}

In order to use the method of 
the amplitude ratio \vs\ phase difference diagram we need a
more complete monitoring of \epscep. This would require a 
multi-site campaign with monitoring in two or more filters. 
This has been done for a number of \dss\ stars where
extensive multi-site campaigns with long temporal coverage
were carried out. 
Examples are XX~Pyx \citep{handler00} and BI~CMi \citep{breger02}.
We propose to use high-resolution time-series spectroscopy 
to study line-profile variations, which will make it possible to identify the modes.
In addition, information on mode splitting (and azimuthal order $m$) can be achieved.
When this is combined with additional photometric 
observations from ground (or space) we may hope to improve on the present work.

Perhaps the most interesting result from the current study is that
we have detected several significant (S/N=4.5--6.5) frequencies with very 
low amplitude: seven frequencies have amplitudes below 0.5 ppt. 
For many years is has been a puzzle why only 
some of the frequencies predicted from models of \dss\ stars were in fact detected.
\citet{breger05} discuss their results for FG~Vir and point out that
the ``missing modes'' are indeed there but that the detection level in
previous studies was too poor.
About two thirds of the frequencies detected in FG~Vir have amplitudes
in $y$ below 0.5 ppt. From the \wire\ data of \epscep\ we also
find that several frequencies have amplitudes below 0.5 ppt ($\simeq y$ filter). 
The reason why we do not find even more frequencies with low amplitude 
is most likely that the frequency resolution in the data is too low.
Even so, our results give support to the suggestion by \citet{breger05} that
the modes predicted by models of \dss\ stars are indeed present in the stars
but one must have long temporal coverage 
with high photometric precision to be able to detect them.

The Canadian \most\ satellite \citep{walker03} 
can monitor stars for up to 60 days with
a duty cycle close to 100\% with high photometric precision.
Preliminary results for a \dss\ star observed as a secondary target by \most\
show about 80 frequencies to be present with amplitudes as low 
as 0.1~mmag (J.~M.\ Matthews, private communication).
In the coming years two more dedicated photometry missions 
will be launched: \corot\ and \kepler.
The \corot\ mission \citep{baglin01} will monitor several relatively bright
stars for up to 150 days to obtain photometry with very high precision and
with a duty cycle close to 100\%.
However, observations are done in one filter 
only for all the missions mentioned here: \wire, \most, \corot, and \kepler.
Hence, ground-based support observations with 
multiple filters and/or spectroscopic measurements 
are needed in order to be able to identify the modes.

The present analysis of \epscep\ has shown that much more 
complete and carefully planned ground-based observations are needed
to avoid problems resulting from a complex spectral window.
Also, it is necessary to collect enough multi-colour photometry
to be able to measure phases 
and amplitudes with the required accuracy to be able to identify the modes.
Multi-site campaigns that overlap in time with the space-based 
observations should be arranged for the future missions. 
For example, most primary \corot\ targets are quite bright ($V\simeq6$) and
so long-term (\ie\ months) monitoring with small 20--80\,cm class telescopes 
with \str\ filters and a stable photometer will be adequate to 
collect the necessary data.
The missions have several secondary targets which are monitored
at the same time as the primary target. Thus it will be a 
difficult but potentially valuable task
to coordinate and collect all the necessary data.

\begin{acknowledgements}

HB is supported by the Danish Research Agency 
({Forskningsr\aa det for Natur og Univers}), 
the Instrument center for Danish Astrophysics (IDA),
and the Australian Research Council.
HB is grateful to Torben Arentoft and Gerald Handler for useful discussions. 
JCS acknowledges support from the Instituto de 
Astrof\'{\i}sica de Andaluc\'{\i}a through an I3P contract
financed by the European Social Fund and from the Spanish 
Plan Nacional del Espacio under project ESP2004-03855-C03-01.

\end{acknowledgements}


\begin{thebibliography}{}  


\bibitem[Aerts \etal, 2002]{aerts02} Aerts, C., Handler, G., 
 Arentoft, T., Vandenbussche, B., Medupe, R., \& Sterken, C.\ 
  2002, \mnras, 333, L35 

\bibitem[Baade \etal, 1993]{baade93} Baade, D., Bardelli, S., 
 Beaulieu, J.~P., \& Vogel, S.\ 
 1993, \aap, 269, 195 
 
 \bibitem[Baglin \etal, 2001]{baglin01} Baglin, A., Auvergne, 
   M., Catala, C., Michel, E., \& \corot\ Team 
 2001, ESA SP-464: SOHO 10/GONG 2000 Workshop: 
 Helio- and Asteroseismology at the Dawn of the Millennium, 10, 395 

\bibitem[Bernacca \& Perniotto, 1970]{bernacca70} Bernacca, P.L., Perniotto, M.\
 1970, Contr.\ Oss.\ Astrof.\ Padova in Asiago, 239, 1
 
\bibitem[Bessell \etal, 1998]{bessell} Bessell, M.~S., 
 Castelli, F., \& Plez, B.\ 
 1998, \aap, 333, 231 

\bibitem[Breger, 1966]{breger66} Breger, M.\ 
 1966, \apj, 146, 958 

\bibitem[Breger \etal, 2002]{breger02} Breger, M. \etal\ 
 2002, \mnras, 329, 531 

\bibitem[Breger \etal, 2005]{breger05} Breger, M. \etal\ 
 2005, \aap, 435, 955

\bibitem[Breger \& Pamyatnykh(2006)]{breger06} Breger, M. \& Pamyatnykh, A.~A.\ 
 2006, \mnras, 339 
 
 \bibitem[Buzasi \etal, 2000]{buzasi00}
 Buzasi, D. L., Catanzarite, J., Laher, R.\ \etal\
 2000, ApJ, 532, 133 

\bibitem[Buzasi \etal, 2005]{buzasi05}
 Buzasi, D. L., Bruntt, H., Bedding, T. R.\ \etal\
 2005, ApJ, 619, 1072 

\bibitem[Bruntt \etal, 2005]{bruntt05}
 Bruntt, H., Kjeldsen, H., Buzasi, D.L., Bedding, T.\ R.
 2005, ApJ, 633, 440

\bibitem[Bruntt \& Buzasi(2006)]{bruntt06} Bruntt, H., \& Buzasi, D.~L.\ 
 2006, Memorie della Societa Astronomica Italiana, 77, 278 

\bibitem[Costa \etal, 2003]{costa03} Costa, V., Rolland, A., 
 L{\' o}pez de Coca, P., Olivares, I., \& Garc{\'{\i}}a-Pelayo, J.~M.\ 
 2003, Asteroseismology Across the HR Diagram, 397 

\bibitem[\protect\citeauthoryear{Daszy{\'n}ska-Daszkiewicz, Dziembowski, \& 
Pamyatnykh}{2003}]{dasz03} Daszy{\'n}ska-Daszkiewicz J., 
Dziembowski W.~A., Pamyatnykh A.~A., 2003, A\&A, 407, 999 

\bibitem[Erspamer \& North, 2002]{erspamer02} Erspamer, D.\ \& 
 North, P.\ 2002, \aap, 383, 227 

\bibitem[Erspamer \& North, 2003]{erspamer03} Erspamer, D.\ \& North, P.\ 
 2003, \aap, 398, 1121 

\bibitem[ESA, 1997]{hipparcos97} ESA 
 1997, The Hipparcos and Tycho Catalogues, ESA SP-1200

\bibitem[Fesen, 1973]{fesen73} Fesen, R.~A.\ 
 1973, \pasp, 85, 732 

\bibitem[Garrido \etal, 1990]{garrido90} Garrido, R., 
Garcia-Lobo, E., \& Rodriguez, E.\ 
 1990, \aap, 234, 262 

\bibitem[Gray, 1971]{gray71} Gray, D.~F.\ 
 1971, \pasp, 83, 103 

\bibitem[Hacking \etal, 1999]{hacking99}
 Hacking, P., Lonsdale, C., Gautier, T.\ \etal\ 
 1999, ASP Conf.\ Ser., 177, 409, Eds.\ M.D.\ Bicay, R.M.\ Cutri, \& B.F.\ Madore

\bibitem[Handler \etal, 1997]{handler97} Handler, G.\ \etal\
 1997, \mnras, 286, 303 
  
\bibitem[Handler \etal, 2000]{handler00} Handler, G.\ \etal\ 
 2000, \mnras, 318, 511 
 
\bibitem[Hauck \& Mermilliod, 1998]{hauck98} 
   Hauck, B., \& Mermilliod, M.,
 1998, A\&AS, 129, 431

\bibitem[Kennelly \etal, 1999]{kennelly99} Kennelly, E. J., Brown, T. M., 
 Ehrenfreund, P., Foing, B., Hao, J., Horner, S., Korzennik, S., 
 Nisenson, P., Noyes, R., Sonnentrucker, P.
 1999, ASPC, 185, 264

\bibitem[Kjeldsen \& Frandsen(1992)]{kjeldsen92} Kjeldsen, H., \& Frandsen, S.\ 
 1992, \pasp, 104, 413 

\bibitem[Kupka \& Bruntt, 2001]{kupka01} Kupka, F., \& Bruntt, H.\ 
 2001, First \corot/\most/\most\  Ground  Support Workshop, 39 
 
\bibitem[Lejeune \& Schaerer, 2001]{lejeune} Lejeune, T.\ \& Schaerer, D.\ 
 2001, \aap, 366, 538 
 
\bibitem[Lenz \& Breger, 2005]{lenz05} Lenz, P., \& Breger, M.\ 
 2005, CoAst, 146, 53 

\bibitem[Loumos \& Deeming, 1978]{loumos78} Loumos, G.~L., \& 
 Deeming, T.~J.\ 
 1978, \apss, 56, 285 
 
\bibitem[Mantegazza \& Poretti(2002)]{mante02} Mantegazza, L., 
\& Poretti, E.\ 2002, \aap, 396, 911 

\bibitem[Mittermayer \& Weiss(2003)]{mitter03} Mittermayer, P., 
\& Weiss, W.~W.\ 2003, \aap, 407, 1097 

\bibitem[Montgomery \& O'Donoghue, 1999]{montgomery} 
         Montgomery, M.H. \& O'Donoghue, D.
 1999, Delta Scuti Star Newsletter, 13, 28
 
\bibitem[Moya \etal, 2004]{moya04} Moya, A., 
 Garrido, R.\ \& Dupret, M.~A.\ 
 2004, \aap, 414, 1081 

 \bibitem[Rodr{\'{\i}}guez \etal, 2000]{rodriguez00} 
 Rodr{\'{\i}}guez, E., L{\' o}pez-Gonz{\' a}lez, M.~J., \& L{\' o}pez de Coca, P.\ 
 2000, \aaps, 144, 469 

\bibitem[Rogers, 1995]{rogers95} Rogers, N.Y.\ 
 1995, CoAst, 78

\bibitem[Royer \etal, 2002]{royer02} Royer, F., Grenier, S.\
 Baylac, M.-O., G\'omez, A.E., \& Zorec, J.,
 2002, A\&A, 393, 897 

\bibitem[Schwarzenberg-Czerny, 1991]{schwarzenberg91} 
 Schwarzenberg-Czerny, A.\ 
 1991, \mnras, 253, 198 
 
\bibitem[Suarez \etal, 2005]{suarez05}
 Su\'arez, J.C., Bruntt, H., Buzasi, D.\
 2005, A\&A, 438, 633

\bibitem[Walker \etal, 2003]{walker03} Walker, G., Matthews, J., Kusching, R.\ \etal\
 2003, \pasp, 115, 1023

\bibitem[Zima, 1997]{zima97} Zima, W.\ 
 1997, Delta Scuti Star Newsletter, 11, 37 

\bibitem[Zima \etal, 2002]{zima02} Zima, W.\ \etal\
 2002, 
  ASP Conf.~Ser.~259: IAU Colloq.~185: Radial and Non-radial Pulsations as 
  Probes of Stellar Physics, 259, 598 

\end{thebibliography}
\end{document}